\documentclass[twocolumn,showpacs,preprintnumbers,amsmath,amssymb]{revtex4}

\newcounter{ctr}

\usepackage{graphicx}
\usepackage{epsfig}
\usepackage{dcolumn}
\usepackage{bm}

\begin{document}

\title{Dynamics of Coupled Adaptive Elements : Bursting and Intermittent Oscillations Generated by Frustration in Networks }

\author{Masayo Inoue$^1$ and Kunihiko Kaneko$^{1,2}$}
\affiliation{
$^1$Department of Basic Science, Graduate School of Arts and Sciences, University of Tokyo, \\
3-8-1 Komaba, Meguro-ku, Tokyo 153-8902, Japan \\
$^2$ ERATO Complex Systems Biology Project, JST,  \\
3-8-1 Komaba, Meguro-ku, Tokyo 153-8902, Japan \\ }
\date{\today}

\begin{abstract}
Adaptation to environmental change is a common property of biological systems. Cells initially respond to external changes in the environment, but after some time, they regain their original state. By considering an element consisting of two variables that show such adaptation dynamics, we studied a coupled dynamical system containing such elements to examine the diverse dynamics in the system and classified the behaviors on the basis of the network structure that determined the interaction among elements. For a system with two elements, two types of behaviors, perfect adaptation and simple oscillation, were observed. For a system with three elements, in addition to these two types, novel types of dynamics, namely, rapid burst-type oscillation and a slow cycle, were discovered; depending on the initial conditions, these novel types of dynamics coexisted. These behaviors are a result of the characteristic dynamics of each element, i.e., fast response and slow adaptation processes. The behaviors depend on the network structure (in specific, a combination of positive or negative feedback among elements). Cooperativity among elements due to a positive feedback loop leads to simple oscillation, whereas frustration involving alternating positive and negative interactions among elements leads to the coexistence of rapid bursting oscillation and a slow cycle. These behaviors are classified on the basis of the frustration indices defined by the network structure. The period of the slow cycle is much longer than the original adaptation time scale, while the burst-type oscillation is a continued response that does not involve any adaptation. We briefly discuss the universal applicability of our results to a network of a larger number of elements and their possible relevance to biological systems. 
\end{abstract}

\maketitle

\begin{center}
\textbf{I. INTRODUCTION}
\end{center}

Adaptation is a common property of biological organisms.  Individual organisms initially sense and respond to external changes in their environment; however, after some time, they become habituated to the  new environment, and hence, the response disappears. Adaptation can occur on a time scale of generations, but it also progresses on much shorter time scales of seconds, minutes, or hours, i.e., within the lifetime of an individual organism. Here, we focus on the adaptive behavior on such short timescales; such behavior is generally observed not only in multicellular organisms but also in unicellular organisms like \textit{E.coli} or \textit{paramecium} and is regarded as a general property of a cell.

The study of cellular adaptation behavior has been pioneered by Koshland\cite{Koshland_science77, Koshland_science82} and Oosawa\cite{Oosawa_JTB77, Oosawa_exp}. Koshland classified adaptation as absolute(perfect) and partial adaptation. Environmental change has an impact on intracellular states such as gene expression patterns through reaction networks, for example, by signal transduction. In absolute adaptation, variables representing the concentration of some chemicals start to change in response to environmental change, and then they regain their original values even though the environmental change is not reversed. Indeed, several examples of simple biochemical circuits have been proposed to represent such adaptive behaviors\cite{Goldbeter, AsakuraHonda_JMB84, Leibler_CRAS01}. In addition to theoretical interest, experimental verification of such adaptation processes has gained much attention\cite{Block_JB83, Karatan_JBC01, Gasch_MBC00, Shay_MSB07}.

Adaptive response is not limited to a specific variable of a biological system. Not one but many variables regain their original values within a certain time after an environmental change has occurred. This type of response has been studied in the microarray analysis of gene expression levels\cite{Gasch_MBC00, Shay_MSB07}. Often, some partial units in a whole biological system have a tendency to maintain their state, independent of the external conditions $-$ this is known as homeostasis. On the other hand, they also have to respond to an external change in order to survive.  Hence, the adaptive behavior is expected to be an inherent process that involves many partial units of a biological system. Then, the dynamic behavior of the whole system is determined by the behavior of several adaptation units that interact with each other. 

Now, in addition to revealing the processes and mechanisms in an adaptation unit, i.e., an adaptation module, it is also important to study a system consisting of several adaptation units that are not isolated but interact with other units. Here, such a unit of adaptive behavior can be a cell in a colony of interacting cells or in a multiceullar organism. In this case, the interaction occurs through chemical signals resulting in cell-cell communication. Alternatively, a certain part of a global intracellular reaction network can function as a unit to exhibit adaptive behavior. In this case, each unit interacts through reaction paths in the network. For example, a signal transduction network or metabolic network is highly complicated, involving many feedback and feedforward interactions or multi step reactions. Each adaptive or non-adaptive unit is embeded within a cell and receives inputs not only from outside the cell but also from other units within the intracellular network. Such interaction between units should play an important role in the global function of a cell.

In this study, we examine the behavior of a coupled system consisting of elements that exhibit adaptive behavior. If each element is a module in a chemical reaction network or a reaction motif, the coupled adaptive system represents a global intracellular reaction network. If each element represents a cell, the system represents the cooperative behavior of cells within a multicellular system.

As for dynamical systems with many degrees of freedom, coupled oscillators\cite{Kuramoto} and coupled chaotic systems (such as coupled maps)\cite{KanekoTsuda, Kaneko_PhysD90} have  been extensively studied over decades. Several novel concepts such as synchronization transition, clustering chaotic itinerancy, and collective chaos have been developed in these studies. In contrast, systems of coupled adaptive elements have not been systematically studied thus far. However, such a study may help develop novel concepts and reveal nontrivial interesting behaviors. In fact, dynamics that facilitate perfect adaptation have the following characteristics that distinguish them from other systems: when a specific parameter of the system is changed, the response to this change appears first as a change in the values of the variables, but later, some variables regain their original values independent of the parameter values. This independence results in the imposition of some constraints. Further, the above behavior implies the existence of two time scales: one for response and the other for relaxation to the original value. 
Therefore, the manner of return to the original values and the existence of the two time scales result in nontrivial dynamics in a coupled system. Here, we classify the  behaviors of coupled adaptive systems on the basis of the interaction network structure among elements. For the classification, we will introduce the measure of frustration to characterize the interaction among elements. 

The paper is organized as follows. In section II, we introduce a model of coupled adaptive elements. In section III, we present results for the two-element case. The three-element case is studied in section IV. In this case, the coexistence of a slow cycle and fast bursting oscillations is observed when frustration is taken into account. Extensions to cases with a larger number of elements will also be discussed.

\begin{center}
\textbf{II. MODEL}
\end{center}

Here, we introduce a model of coupled adaptive elements. In this model, elements that show adaptive behavior to an input interact with each other.  Here, adaptation refers to dynamics by which some variables regain their original value, independent of the value of the external inputs.

Now, we consider a minimal system of such adaptation. When the value of a constant external input $S$ is changed, first, a state variable $u$ changes in response to this input, however, after some time, $u$ regains its original value independent of the value of $S$. This system is a type of excitable system in which a state variable changes in response to a stimulus and then returns to the pre-stimulated state. The simplest way to construct such dynamics is to introduce another variable ($v$) that "absorbs" the changes in $S$. Hence, we consider the following two-variable dynamical system:
\begin{eqnarray}
\frac{du}{dt}=f(u, v ; S) , \nonumber \\
\frac{dv}{dt}=g(u, v ; S), 
\end{eqnarray}
where the fixed point solution $u^* , v^*$ given by $f(u^*, v^* ; S)=0,\ g(u^*, v^* ; S)=0 $ should be set to be stable. The above condition is satisfied if $u^*$ is invariant against the change in $S$, even though  $f$ increases with $S$. Then, the postulate for the response to $S$ and adaptation is satisfied because $u$ first increases with $S$ and then regains the value $u^*$. 

\begin{figure}
\begin{center}
\scalebox{0.35}{\includegraphics{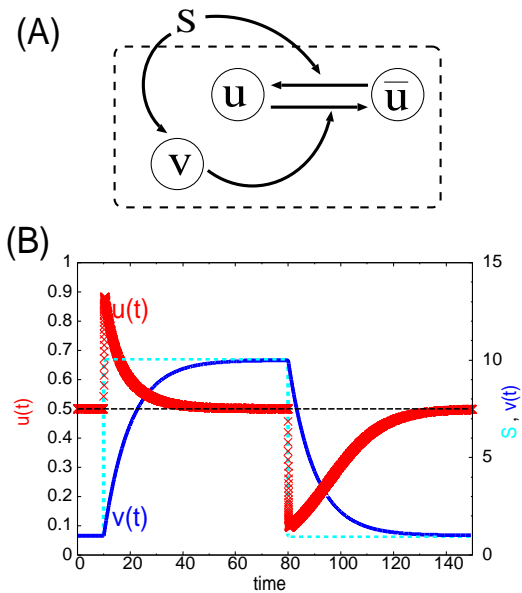}}
\caption{(Color online) (A) schematic of a reaction system represented by eq.(\ref{eq_adapt}). The area enclosed by the dashed line contains one element. (B) A typical example of a response that eq.(\ref{eq_adapt}) obeys; $u$ (red $\times$ point, plotted on the vertical axis to the left) represents adaptive dynamics and $v$ (blue solid line, plotted on the vertical axis to the right) represents relaxation according to the change in $S$ (cyan dotted line, plotted on the vertical axis to the right). 
              }
    \label{fig:element}
\end{center}
\end{figure}

As an example of such adaptive dynamics, we consider a chemical reaction. In the reaction process, time evolution is governed by a set of rate equations in which variables $u$ and $v$ represent chemical concentrations. Indeed, there have been several models that exhibit adaptive dynamics\cite{MIKK_PRE06, AsakuraHonda_JMB84, Leibler_CRAS01, ErbanOthmer_SIAM04}. In this article, we adopt a system obtained by adiabatically eliminating a variable from a model originally introduced by Levchenko and Iqlesias\cite{Levchenko}. 
In this simplified model, a chemical $U$ is either active or inactive:  $u$ represents the concentration of the chemical $U$ in the active state and $\overline{u}$ represents that in the inactive state. We set the sum of $u$ and $\overline{u}$ to $1$ ($u + \overline{u} = 1 $). An external input ($S$) activates $U$ by catalyzing the process from $\overline{u}$ to $u$, while it is also involved in the synthesis of another chemical $V$, which acts as a catalyst to deactivate $U$. When the degradation of $V$ is also included (Fig.\ref{fig:element}), the kinetics is represented as
\begin{eqnarray}
\frac{du}{dt} =\frac{S (1- u) -u v}{\eta} , \ \ \  \frac{dv}{dt} =\frac{S- v}{\mu}.
 \label{eq_adapt}
\end{eqnarray}
We rescale time as $\tilde{t} = t / \eta$ in all the following discussions, and by setting $\tilde{\mu} = \mu / \eta$, eq.(\ref{eq_adapt}) is re-written as
\begin{eqnarray}
\frac{du}{d\tilde{t}} =S (1- u) -u v , \ \ \  \frac{dv}{d\tilde{t}} =\frac{S- v}{\tilde{\mu}}
 \label{eq_adapt_re}
\end{eqnarray}

This equation has a stable fixed point $v^* = S$, $u^* = {\overline{u}}^* =0.5$, which is independent of $S$. When the input $S$ increases (decreases), the concentration of $U$ in the active (inactive) state increases; therefore, $u$ increases (decreases) from its steady state value ($u^* = {\overline{u}}^* =0.5$) according to the change in $S$, but after some time, it regains the original value $u^*$ as $v$ relaxes to the value $S$. 
The dynamics represented by eq.(\ref{eq_adapt_re}) have two time scales. The time scales are defined by the reciprocals of the two eigenvalues of the dynamics. The eigenvalues are calculated by linear stability analysis around the steady state. The first time scale is given by  $\tau_s = 1/ (2S)$, which corresponds to the response time to the change in $S$, and the second is given by $\tau_a = \tilde{\mu}$, which is the relaxation time to the original value (i.e., adaptation). Here, we postulate faster response and slower adaptation so that $\tau_s = 1 / (2S) < \tau_a = \tilde{\mu}$.  See \cite{Oosawa_JTB77, MIKK_PRE06} for the relevance of this condition to biological adaptation.

Each element in our coupled system follows the above reaction mechanism that involves two chemicals ($u$, $v$) and proceeds according to eq.(\ref{eq_adapt_re}). Now, we consider interactions between such elements $j$ and $i$. We assume that the interaction between these elements can be modeled using the input term $S$, i.e., an output $u_j$ from an element \textit{j} triggers an input to some other element \textit{i} ($S_i$), and thus, the process $u_1, u_2,\cdots \rightarrow S_i$ exists; this process is represented by
\begin{eqnarray}
S_i = h(u_1, u_2, \cdots ) .
\end{eqnarray}

If each element $i$ represents a cell, this interaction represents cell-cell communication through signal molecules, whereas if each element $i$ represents a set of chemical components in an intracellular reaction, the interaction represents signal transduction in an intracellular reaction network.  To choose a specific type of interaction, we take the following two facts into account. First, in a biological reaction, input-output relations often take a sigmoidal form with some threshold, as in an enzyme reaction in which relations are modeled by a Michaelis-Menten type or Hill function. Further, cell-cell interactions can often be represented by such threshold functions, e.g., in binding at receptors or in electric interaction in nerve cells. Hence, we adopt a such sigmoidal reaction to represent interactions between elements. The input $S$ is transmitted according to the concentration of $u$ in the active state. When the balanced state ($u = {\overline{u}} = u^* (=0.5) $) is set to the threshold, the signal is transmitted depending on the sign of $u- u^*$. Hence, each element receives  a signal as a function of $\sum_j C_{ij} \left[ \tanh \{ \kappa (u_j (t) - u^* )  \} \right]$, where $C_{ij}$ represents an interaction factor from element $j$ to $i$, and the element of the interaction matrix $C_{ij}$ is $0$, $1$, or $-1$ depending on whether the signal is non-existant, excitatory, or inhibitory. We exclude self-catalysis and define $C_{ii} =0$. 
Second, catalytic reactions often occur in succession to relay signal transduction, and therefore, the effects of the components are cascaded. Note also that the change in the signal concentration often occurs on the "logarithmic scale".  Without loss of generality, we can set $S=1$ when there is no input from other elements, $S=S_{max}>1$ if the received signal is excitatory, and $S=S_{min}=S_{max}^{-1}$ if it is inhibitory; thus, we obtain 
\begin{eqnarray}
S_i = \exp \left[ \frac{\sigma}{N_{i}} \sum_j C_{ij} \Big\{ \tanh \{ \kappa (u_j (t) - u^* ) \} \Big\} \right] ,
 \label{eq_exp_interaction}
\end{eqnarray}
where $N_{i}$ is the number of inputs that the element $i$ receives, i.e., $N_i= \sum_{j=1}^{N} | C_{ij} | $, and thus, comparison with cases involving  different networks with a different number of paths is straightforward. With this form, $S_{max}$ is given by the coupling strength $\sigma$ as $S_{max} = e^{\sigma}$. 
Without loss of generality, we can set $S_{max}=10$ and $\sigma = \ln 10$ in order to limit the range of $S$ to  $10^{-1} - 10^{+1}$; then, the form of the interaction is 
\begin{eqnarray}
S_i = 10^{ \frac{1}{N_{i}}\sum_j C_{ij} \left[ \tanh \{ \kappa (u_j (t) - u^* )  \} \right] } .
 \label{eq_interaction}
\end{eqnarray}
We use $\kappa = 50$ unless otherwise mentioned.  The response time scale of each element is given by $\tau_s = 1 / (2S)$ and that for adaptation by $\tau_a = \tilde{\mu}$. Setting $\tilde{\mu}=10$, we can use $\tau_s = 0.05 - 5$ and $\tau_a = 10$ in order to satisfy the condition for fast response and slow adaptation ($\tau_s < \tau_a $). There is no qualitative change in our results as long as the threshold-type interaction is chosen and $\tau_s < \tau_a $ irrespective of the parameter value. Even if the exponential coupling form as in eq.(\ref{eq_exp_interaction}) is not adopted, most of the behaviors to be reported here are obtained.

\begin{figure}
\begin{center}
\scalebox{0.35}{\includegraphics{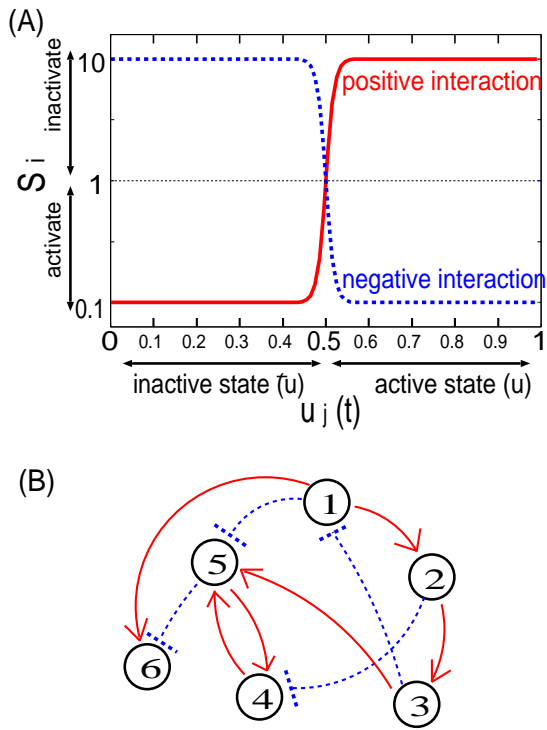}}
\caption{(Color online) (A) interaction function represented by eq.(\ref{eq_interaction}). The red solid line represents a positive interaction and the blue dotted line represents a negative interaction. (B) conceptual scheme of a coupled system. Each numbered circle represents an element with $(u,v)$ adaptive dynamics. The bold red and blue dotted arrows represent positive and negative interactions, respectively.
              }
    \label{fig:interaction}
\end{center}
\end{figure}

In the present model, each element communicates with other elements according to its state, i.e., depending on whether it is active ($u > u^*$) or inactive ($u < u^*$). If $C_{ij}>0$, an active (inactive) element $j$ activates (inhibits) the following element $i$, respectively, and the converse is true  for $C_{ij} <0$ (Fig.\ref{fig:interaction}). We use the above normalization (summation $N_{i}$) so that elements receiving a positive signal have identical input values and identical response time scales, whereas those receiving a negative signal have smaller input values (equal for all such signals) corresponding to a longer time scale.

\begin{center}
\textbf{III. RESULTS FOR TWO-ELEMENT CASE}
\end{center}

\begin{figure}
\begin{center}
\scalebox{0.5}{\includegraphics{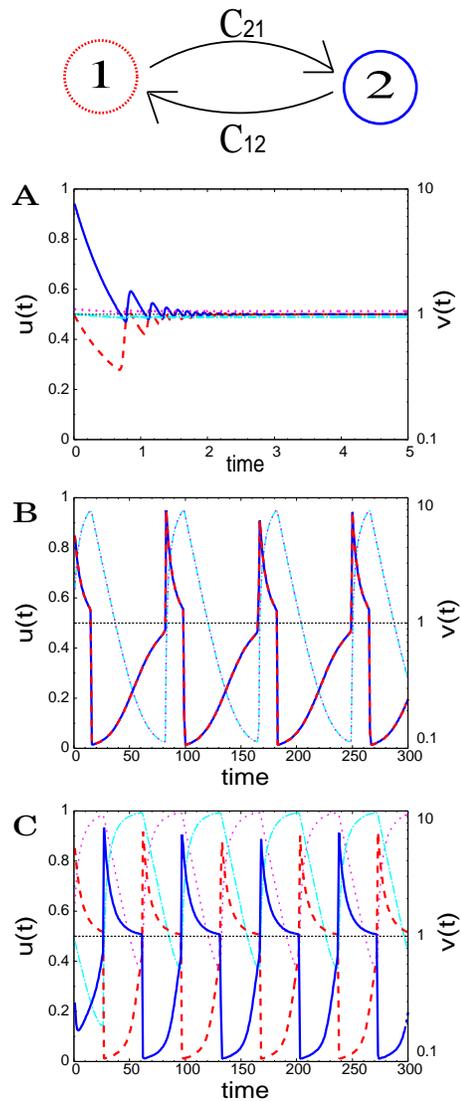}}
\caption{(Color online) Time series of $u$ and $v$ for a two-element system. $\mathbf{A}$ ($C_{12} \times C_{21} < 0$) corresponds to perfect 
adaptation behavior. $\mathbf{B}$ ($C_{12} > 0, C_{21} > 0$) and $\mathbf{C}$ ($C_{12} < 0, C_{21} < 0$) correspond to oscillations in phase and in anti phase, respectively. In each picture, the red dashed line shows the dynamics of $u_1$; the blue solid line, the dynamics of $u_2$; the magenta dotted line, the dynamics of $v_1$; and the cyan chained line, the dynamics of $v_2$. 
               }
    \label{fig:2dynamics}
\end{center}
\end{figure}

First, we present results for the case involving two elements ($1$ and $2$). When one of the $C_{ij} $ values was $0$, the system simply converged to the fixed point with $u^*=0.5$ and $v^*=1$. When one of the elements does not receive any signal, it returns to the original fixed point, and then, the other element does so as well. In addition to such a trivial case, there are three cases for different $C_{12} $ and $C_{21} $: positive-positive, positive-negative, and negative-negative interactions. 
Depending on the type of interaction, two types of behaviors, perfect adaptation and oscillation, are observed (Fig.\ref{fig:2dynamics}). In perfect adaptation, each element eventually shows perfect adaptation, and therefore, the overall system also adapts perfectly and regains the original steady state. 
In the case of oscillation, each element cannot adapt to the steady state and repeats the cycle of response and adaptation. The output of element $1$ causes a change in the input of element $2$ according to the interaction function (\ref{eq_interaction}), and element $2$ shows response and adaptation. This output of element $2$ in turn causes a change in $S_1$. Thus, response and adaptation are repeated continuously. 

If one interaction coefficient is positive and the other is negative ($C_{12} C_{21} < 0$), the system exhibits perfect adaptation. 
Without loss of generality, we choose  $C_{21} > 0$ and $C_{12} < 0$. When element $1$ is activated ($u_1 > u^*$), element $2$ is also activated by positive interaction ($C_{21} > 0$). On the other hand, negative interaction with element $2$ ($C_{12} < 0$) inactivates element $1$. Since the inputs received by the elements have opposite signs, excitatory output is not sustained. The system then undergoes damped oscillations and eventually exhibits perfect adaptation. 
In contrast, the system shows a sustained oscillation if $C_{12} C_{21} > 0$. Depending on the sign of $C_{12}$, elements $1$ and $2$ oscillate in phase (if both $C_{12} $ and $C_{21} $ are positive) or in anti phase (if both $C_{12} $ and $C_{21} $ are negative) between the active and inactive states. The period of the limit cycle is $\sim \tau_a$.

To understand the different behaviors for different networks, we performed linear stability analysis around the steady state ($u_i = u^* = 0.5$ and $v_i = v^* = 1 $). 
Setting $u_i = u^* + \delta u_i$ and $v_i = v^* + \delta v_i$, the linearlized form of eq.(\ref{eq_adapt_re}) is given by 
\begin{align}
\frac{d \delta u_{i}}{d\tilde{t}} &= -2 \delta u_{i} - \frac{1}{2} \delta v_{i} + \frac{\kappa \sigma}{2} \sum_{j} C_{ij} \delta u_{j} , \nonumber \\
\frac{d \delta v_{i}}{d\tilde{t}} &= - \frac{1}{\tilde{\mu}} \delta v_{i} + \frac{\kappa \sigma}{2 \tilde{\mu}} \sum_{j} C_{ij} \delta u_{j} .
 \label{eq_linear}
\end{align}

Defining $\alpha = (\frac{\kappa \sigma}{2} )^2 C_{12} C_{21} $, the four eigenvalues $\lambda$ in the two-element case are given as solutions of the characteristic equation
\begin{eqnarray}
(\lambda + \frac{1}{\tilde{\mu}})^2 (\lambda + 2)^2 - \alpha \lambda^2 = 0 .
 \label{eq_2eigen}
\end{eqnarray}
When $C_{12} C_{21} < 0$, as expected, the real parts of all four eigenvalues are negative, and thus, the original steady state is stable and perfect adaptation is guaranteed. In contrast to a single-element case, these eigenvalues are complex, which leads to damped oscillation.  When $C_{12} C_{21} > 0$, two real eigenvalues are positive and the other two are negative, implying that the original steady state is an unstable focus, leading to a limit cycle attractor.

\begin{center}
\textbf{IV. RESULTS FOR THREE-ELEMENT CASE}
\end{center}

\begin{figure}
\begin{center}
\scalebox{0.7}{\includegraphics{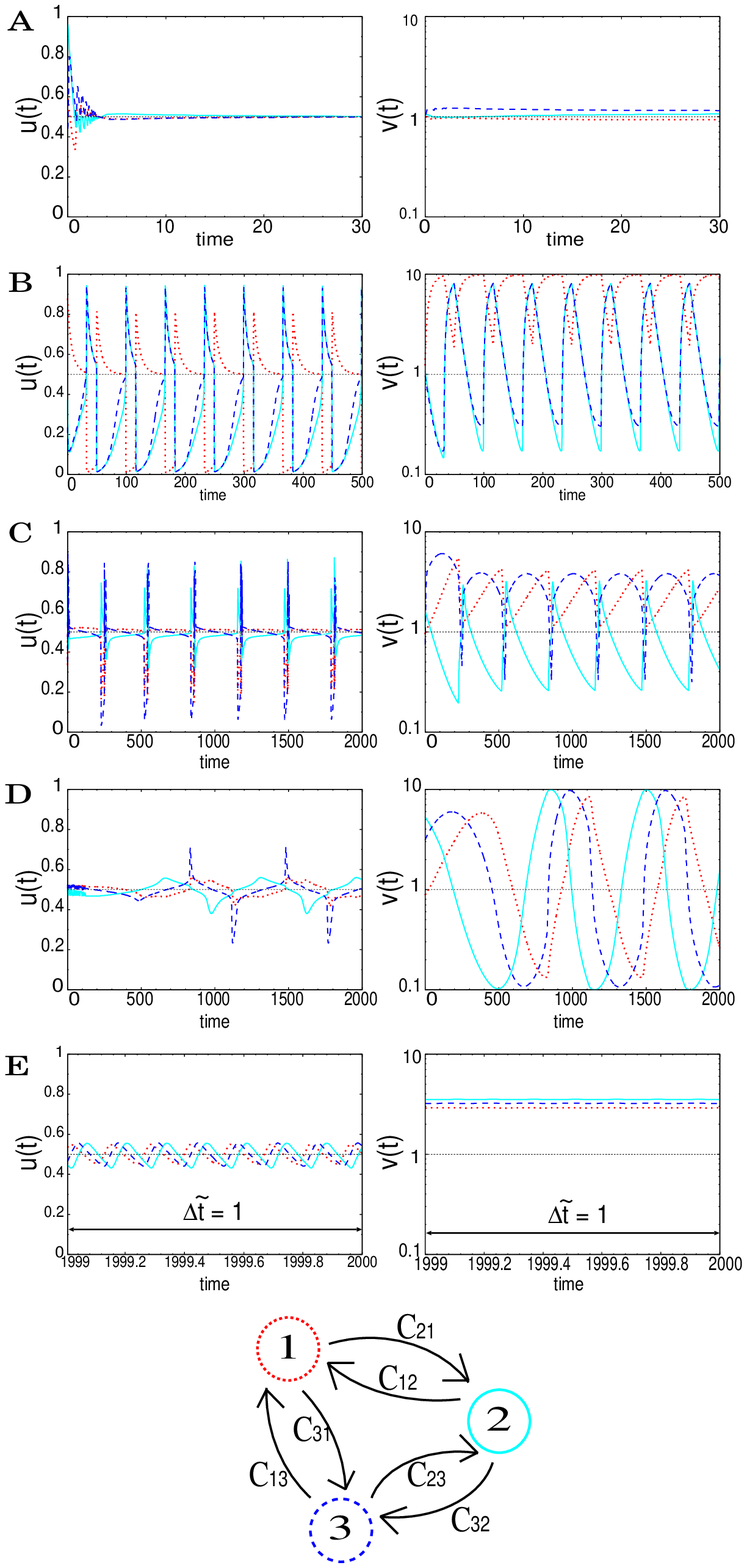}}
\caption{(Color online) Time series of $u$ (left) and $v$ (right) for the three-element system. $\mathbf{A}$ ($C_{12}=-1, C_{13}=-1, C_{21}=1, C_{23}=-1, C_{31}=1, C_{32}=1$) corresponds to perfect adaptation behavior; $\mathbf{B}$ ($C_{12}=-1, C_{13}=0, C_{21}=0, C_{23}=1, C_{31}=-1, C_{32}=1$) corresponds to an example of simple oscillation. $\mathbf{C}$ ($C_{12}=-1, C_{13}=-1, C_{21}=-1, C_{23}=1, C_{31}=1, C_{32}=0$) and $\mathbf{D}$ ($C_{12}=-1, C_{13}=-1, C_{21}=0, C_{23}=1, C_{31}=1, C_{32}=-1$) correspond to a slow cycle and $\mathbf{E}$ (same network as in $\mathbf{D}$) corresponds to a rapid oscillation. Note the difference in the scales of the abscissa for different figures. In each picture, the red dotted line shows the dynamics of element $1$; the cyan solid line, the dynamics of element $2$; and the blue dashed line, the dynamics of element $3$.
              }
    \label{fig:3dynamics}
\end{center}
\end{figure}

\begin{figure}
\begin{center}
\scalebox{0.45}{\includegraphics{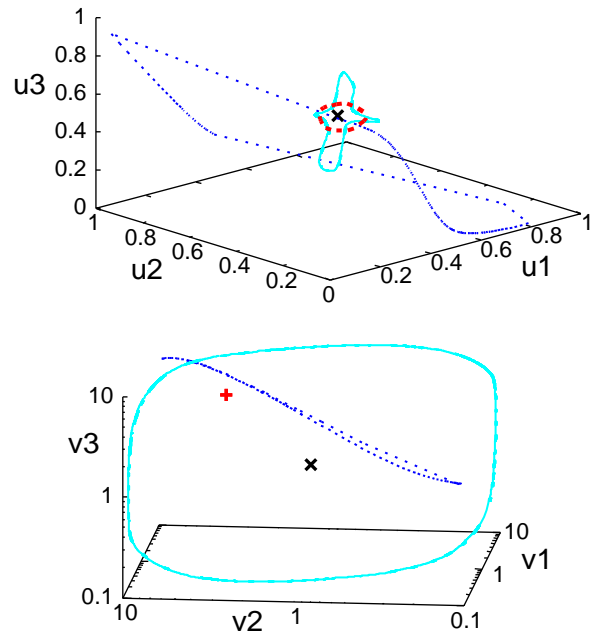}}
\caption{(Color online) Trajectories in the phase space of $u_1,u_2,u_3$ (up) and of $v_1,v_2,v_3$ (down) for three-element systems of type (i) $-$ (iii). Each orbit is drawn after neglecting a sufficiently long transient time. The fixed point (black $\times$ point) represents perfect adaptation and the limit cycle shown by the blue dotted line is an example of simple oscillation. An example of  rapid oscillation is plotted as a red bold broken line for $u_1,u_2,u_3$ and as a $+$ point for $v_1,v_2,v_3$, as the oscillation amplitude in $v$ is negligibly small. Note that this fixed value of the $v$'s differs greatly from the original fixed point that corresponds to perfect adaptation. The cyan solid line for plots of both $u$ and $v$ denotes a slow cycle. The results for (a), (b), (d), and (e) in Fig.\ref{fig:3dynamics} are plotted.
              }
    \label{fig:3orbit}
\end{center}
\end{figure}

For a three-element system,  there are six directed interactions between elements that are given by $C_{ij} $, which takes the value $1, -1$, or $0$. Hence, the total number of possible networks is $3^6 = 729$. By disregarding the cases that are identical due to symmetry and those in which one or two elements are disconnected, we obtain $78$ possible networks. For example, if the third element is driven by only one of the other two elements and does not transmit to other elements, this obviously results in the same behavior as the two-element case, and hence, such networks are excluded. We have studied all of these $78$ cases to find three distinct behaviors: (i) perfect adaptation, (ii) simple oscillation (as in the two-element case), and (iii) coexistence of rapid burst-type oscillation and a slow limit cycle (Fig.\ref{fig:3dynamics}). Among the $78$ systems, $7$ networks show perfect adaptation and $46$ show simple oscillation. These two cases are basically understood as straightforward extensions of the corresponding two-element cases. The remaining $25$ networks show a nontrivial behavior that first appears in the three-element case (Fig.\ref{fig:3orbit}). 

There are two distinct types of dynamics depending on the initial conditions. 
In the rapid burst-type oscillation, each $v_i$ tends toward a fixed value $v_i^{fix}$, with tiny-amplitude oscillations around the value, while $u$ components show a large-amplitude oscillation with a period on the order of $\tau_s$. Note that the values $v_i^{fix}$ are distinct from the original fixed point value $v^*=1$. In the slow cycle, $u$ components approach the values of the original fixed point, remain close to these values for some time, and then depart from it. This process is repeated periodically on a time scale much longer than $\tau_a$ (the adaptation time scale).

Now, we analyze the rapid burst-type oscillation. In this case, the $v_i$'s stay almost constant and no adaptation occurs in $u_i$'s, which oscillate on the order of the fast response time scale. The existence of the rapid burst-type oscillation is associated with the time-scale difference. As $\tilde{\mu}$ tends to infinity, or in other words, as the time scale ratio $\tau_s / \tau_a$ between $u$ and $v$ tends to zero, the influence of $u$ on $v$ through $S$ is averaged out because $u$ oscillates too fast, and hence, the oscillation in $v$ disappears. In fact, we have numerically confirmed that the amplitude of oscillation in the $v$'s vanishes as $\tilde{\mu}$ increases.

\begin{figure}
\begin{center}
\scalebox{0.4}{\includegraphics{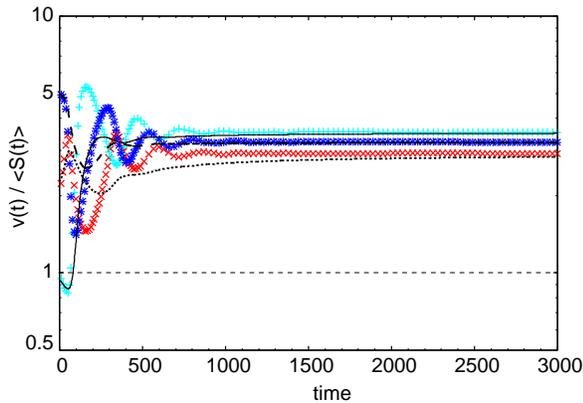}} 
\caption{(Color online) Comparison of the solution $v(t)$ (points) with an approximation solution of $<S>_{\infty}$ obtained from $dv/dt = 0$ (line). The parameters are the same as those in Fig.\ref{fig:3dynamics}(e). The results for elements $1$ (dotted line, red $\times$), $2$ (solid line, cyan $+$), and $3$ (dashed line, blue $\ast$) are shown. Notice that each approximation (line) gives a good solution for the behavior (points).
              }
    \label{fig:limitR}
\end{center}
\end{figure}

Now, we consider this limiting behavior at $\tau _s / \tau _a \sim 1 / \tilde{\mu} \rightarrow 0$.
From the condition $dv / d\tilde{t} = 0$ in eq.(\ref{eq_adapt_re}), ${v_i}^{fix}$ is proved to be a long-time average of $S_i$, i.e., 
\begin{eqnarray}
{v_i}^{fix} = \lim_{T \rightarrow \infty} \frac{1}{T} \int _{0}^{T} S_i (\tilde{t}) d\tilde{t} = <S_i>_{\infty}
 \label{eq_rapidV}
\end{eqnarray}
(Fig.\ref{fig:limitR}). 
Because $dv / d\tilde{t} = 0$, no adaptation occurs and the system is driven only by $u$ variables with the fast response time scale ${\tau_s}' = 1 / (S+v^{fix}) $, which is obtained from
\begin{eqnarray}
\frac{du}{d\tilde{t}} =S (1- u) -u v^{fix} = - (S+ v^{fix}) u + S .
 \label{eq_rapidU}
\end{eqnarray}

Of course, the above solution should be valid only in the limit $\tau_s / \tau_a \rightarrow 0$, but as shown in Fig.\ref{fig:limitR}, the solution agrees well with the present case in which $\tau_s / \tau_a \sim (0.5 - 0.005)$. In this rapid burst-type oscillation, the system enters a state in which the average input from all the other elements is cancelled; consequently, $v$ constantly remains in a state that is away from the original perfect adaptation state.

The rapid burst-type oscillation occurs when successive inputs throughout the loop of three elements change sign, or, in other words, there is "frustration" in the loop interactions (frustration is a term in spin glass theory\cite{spinglass}). 
As a simple example, consider the case in which $C_{21}=1$, $C_{32}=1$, and $C_{13}=-1$. When element $1$ is active, element $2$ is activated through positive interaction with $C_{21}=1$ and element $3$ is also activated through positive interaction with $C_{32}=1$; however, negative interaction with $C_{13}=-1$ inactivates element $1$. Hence, the input inconsistent with its state exists. This argument is true even if we assume that element $1$ is inactive. For any value of $u_i$, such opposite inputs remain, and hence, "frustration" among elements exists, as studied in spin statistical mechanics, e.g., spinglass theory. 
Unlike the case of perfect adaptation, this inconsistency in two-to-one anti-phase relationship cannot be eliminated, and hence, some outputs remain and destabilize the original adapted state. The $v$ variables that cannot change as fast as this opposite input stay almost constant at $v^{fix} = <S>_{\infty}$, whereas only $u$ variables show fast oscillations.

\begin{figure}
\begin{center}
\scalebox{0.4}{\includegraphics{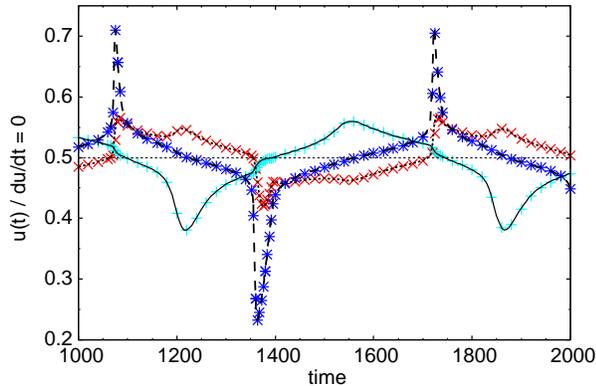}} 
\caption{(Color online) Comparison of the solution $u(t)$ (points) with the following solution obtained from an adiabatic approximation: $u' = S / (v_{fix} + S)$ from $du/d\tilde{t} =0$ (line). The parameters are the same as those in Fig.\ref{fig:3dynamics}(d). The results for elements $1$ (dotted line, red $\times$), $2$ (solid line, cyan $+$), and $3$ (dashed line, blue $\ast$) are shown. Notice that each approximation (line) gives a good solution for the behavior (points).
              }
    \label{fig:limitS}
\end{center}
\end{figure}

Next, we study the slow oscillation in which both $u$ and $v$ components show a periodic change with a time scale much longer than $\tau_a$ (the original adaptation time scale). As $v$ changes slowly, the adiabatic approximation is valid. In particular, this approximation holds under the limit $\tau _s / \tau _a \sim 1 / \tilde{\mu} \rightarrow 0$. 
Slow $v$ variables gradually relax to $S$ values with $\tau_a$, while fast $u$ variables relax to the equilibrium values, defined by $du/d\tilde{t} = 0$, according to the values of the  $v$ variables at each instant within $\tau_s$ (Fig.\ref{fig:limitS}). Thus, $u_i$'s approach the original adapted value and remain close to this value over the time scale $\tau_a$. For a majority of the time, $u_i$'s are adapted to the input $S$ from other elements and show only intermittent, periodic responses, while $v_i$'s change gradually. 
By adiabatically eliminating the variable $u_i$, the motion of slow $v$ variables is obtained. In Fig.\ref{fig:limitS}, we have plotted the present orbit and the solution obtained by this adiabatic elimination; we find that the agreement is rather good. Note that in the present case, the input $S$ varies between $0.1$ and $10$, and indeed, when $S \sim 0.1$, the effective response time $\tau_s = 1 / (2S) \sim \tau_a / 2$. Despite this change in the time scale, the adiabatic approximation agrees well with our simulation results.

\begin{figure}
\begin{center}
\scalebox{0.35}{\includegraphics{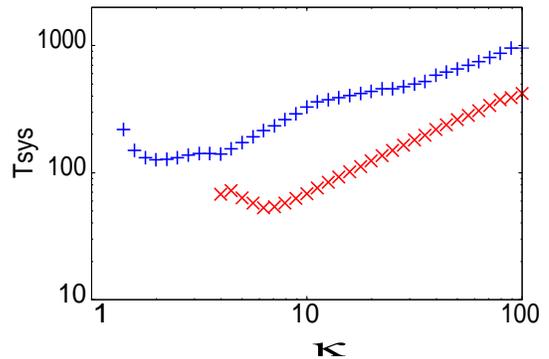}}
\caption{(Color online) Dependence of $T_{sys}$ in the slow cycle on the strength of the interaction function ($\kappa$ in eq.(\ref{eq_interaction})). 
The symbols $\times$ and $+$ represent the results corresponding to networks in Fig.\ref{fig:3dynamics}(c) and Fig.\ref{fig:3dynamics}(d), respectively. As the interaction becomes stronger ($\kappa$ becomes larger), the characteristic time scale gets longer. The time scale of the response ($\tau_s$) is $0.05 - 5$, and the time scale of adaptation ($\tau_a$) is equal to $10$.}
    \label{fig:hill}
\end{center}
\end{figure}

In the slow cycle, the variables $u$ and $v$ undergo a repetitive process in which they approach, remain nearby, and depart from the original fixed point ($u=u^*=0.5, v=v^*=1$); the duration of the slow cycle is much longer than $\tau_a$.
This long time scale is a result of interaction. To verify this claim, we studied the dependence of the period on several parameters in the interaction function (\ref{eq_exp_interaction}). We found that the period depends on the strength of the interaction ($\kappa \sigma /2$ in eq.(\ref{eq_2eigen}) or (\ref{eq_3eigen})). 
Since $\tau_s$ changes with the value of $\sigma$, we study the dependency of the period on the value of $\kappa$. Up to some value of $\kappa$, the interaction function $h$ does not have a sufficiently steep threshold, and hence, the original fixed point is stable and the slow cycle state does not exist. 
Beyond some critical value of $\kappa$, the slow cycle appears as a result of Hopf bifurcation. With a further increase in $\kappa$ or the interaction strength $\kappa \sigma /2$, the period increases even though the adaptation time $\tau_a$ remains fixed (Fig.\ref{fig:hill}). 
The period of burst-type oscillation, in contrast, remains at $\tau_s$ despite the change in $\kappa$.

\begin{figure}
\begin{center}
\scalebox{0.5}{\includegraphics{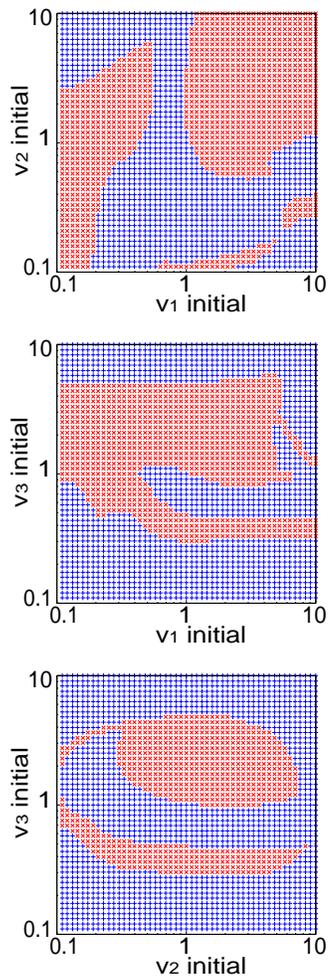}}
\caption{(Color online) Basin structure of rapid oscillation (red $\times$) and slow cycle (blue $+$) for different initial perturbations. The parameters are the same as those in Fig.\ref{fig:3dynamics}(d) and (e). The perturbation is introduced only in each of the indicated variables, and the remaining variables are set to the steady state value ($u^* = 0.5, v^* = 1$). In order to classify the attractors, $T_{sys}$ is computed from the return time after a sufficiently long period that includes transients has been disregarded.
            }
    \label{fig:multi}
\end{center}
\end{figure}

For most networks that we have numerically studied, rapid burst-type oscillation and the slow cycle coexist. There exist only two attractors, depending on the initial conditions (Fig.\ref{fig:multi}). How the two solutions are separated in phase space is clear when we consider the limit $\tau _s / \tau _a \sim 1 / \tilde{\mu} \rightarrow 0$. Under such a limit, the slow cycle is obtained by taking an adiabatic approximation given by $du/d\tilde{t} = 0$, according to the value of $v$ at each instant. On the other hand, rapid burst-type oscillation is obtained in the condition $dv/d\tilde{t} = 0$. The two solutions coexist without canceling each other, since the rapid burst-type oscillation traces around the manifold spanned by $u_i$($i=1,2,3$) whereas the slow cycle traces around the manifold spanned by $v_i$; 
the two oscillations occur in spaces orthogonal to each other. It is remarkable that two attractors whose periods differ by more than two digits coexist in the same network.

\begin{figure}
\begin{center}
\scalebox{0.45}{\includegraphics{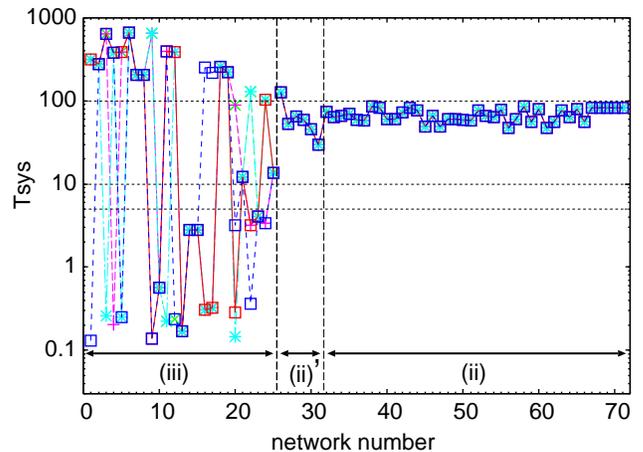}}
\caption{(Color online) Periods $T_{sys} $ of three-element systems (ordinate) plotted against the network number (abscissa). The network numbers are determined arbitrarily but are sorted so that No.$1-25$ correspond to case (iii), where rapid oscillation and 
a slow cycle coexist, and No.$26-71$ correspond to the simple oscillation. The networks that show perfect adaptation are not included here.
Among the networks that shows simple oscillation, No.$26-31$ are networks corresponding to the shaded area in Fig.\ref{fig:phase}. The results for five different initial conditions for each network are overlaid. The values $\tau_s = 0.05 - 5$ and $\tau_a = 10$ are chosen here. The periods are computed from the return time after a sufficiently long period that includes transients has been disregarded.
              }
    \label{fig:Tsys}
\end{center}
\end{figure}

Now, we classify the networks into classes of networks in which one of the following occur: (i) perfect adaptation, (ii) simple oscillation, and (iii) coexistence of rapid burst-type oscillation and a slow cycle; the classification is based on the network structure. 
For cases (ii) and (iii), we have computed the period of the system ($T_{sys}$). The time periods are plotted in Fig.\ref{fig:Tsys} as a function of a network index (abscissa). Since perfect adaptation corresponds to a fixed point solution, networks of type (i) are not included. For the parameter values chosen here, the time scales for a single element are $\tau_s = 1 / (2S) = 0.05 - 5$ for fast response and $\tau_a = \tilde{\mu} = 10$ for slow adaptation. The networks with indices less than $25$ belong to category (iii), where two types of limit cycles coexist. The rapid oscillation has a period $0.05 \leqslant T_{sys} < 5$, which implies that the period is within the range of $\tau_s$. The slow cycle has a period much longer than $\tau_a$: $T_{sys} \gtrsim 10 \tau_a$ for the present parameter values. 
The coexistence of the two solutions on distinctly separated time scales is clearly observable in the figure. The simple oscillation has a period that is slightly larger than, but on the same order as, the adaptation time scale $\tau_a$, i.e., $\tau_a \lesssim T_{sys} \lesssim 10 \tau_a $. For these networks, the simple oscillation is the unique attractor; as seen in the figure, the period of oscillation always lies between the time periods of rapid and slow oscillations. The classification of the type of oscillations on the basis of the order of $T_{sys}$, $\tau_s$, and $\tau_a$ is not influenced by the time scale for each element. 

To classify the cases (i),(ii), and (iii), we again performed a linear stability analysis of eq.(\ref{eq_adapt_re}) around the original, perfectly adapted fixed point solution. The six eigenvalues $\lambda$ are solutions of the following characteristic equation: 
\begin{eqnarray}
(\lambda + \frac{1}{\tilde{\mu}})^3 (\lambda + 2)^3 - \gamma \lambda ^2 (\lambda + \frac{1}{\tilde{\mu}}) (\lambda + 2) - \beta \lambda ^3 = 0 ,
 \label{eq_3eigen}
\end{eqnarray}
where 
\begin{align}
\beta &= (\frac{\kappa \sigma}{2} )^3 (C_{12} C_{23} C_{31} + C_{13} C_{32} C_{21} ) , \\
\gamma &= (\frac{\kappa \sigma}{2} )^2 (C_{12} C_{21} + C_{23} C_{32} + C_{31} C_{13} ) .
 \label{eq_betagamma}
\end{align}

Note that the two indices $\beta$ and $\gamma$ characterize a dominant network structure; $\beta$ indicates whether the system has a positive or negative loop over the three elements. When $\beta > 0$, each element receives an input traversed the entire loop without changing its sign. Inputs to support and suppress activity are received through the loop when an element is active and inactive, respectively. Hence, three elements can behave in unison or in two-to-one anti-phase synchronization. 
If $\beta=0$, the input through the loop is cancelled, while for $\beta < 0$, as previously mentioned, the frustration already exists, and through the loop, each element receives an input inconsistent with its state. The index  $\gamma$, on the other hand, represents the average cooperativity between the two neighboring elements. When $\gamma >0$, two elements can behave in a synchronized manner, while when $\gamma < 0$, they cannot do so because of frustration between the two elements. 

\begin{figure}
\begin{center}
\scalebox{0.55}{\includegraphics{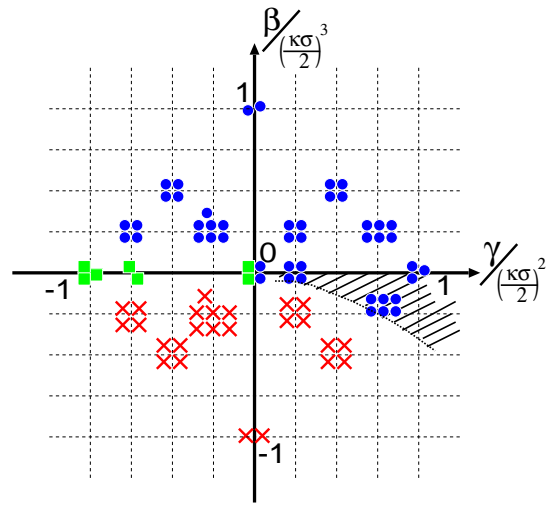}}
\caption{(Color online) Phase diagram of the three behaviors in a three-element system characterized by two indices $\beta$ (ordinate axis) and $\gamma$ (abscissa axis). $\beta$ indicates the consistency around the three elements and $\gamma$ shows the average cooperativity between the neighboring two elements. The definitions of $\beta$ and $\gamma$ are given in the text. The green box represents a network that exhibits perfect adaptation behavior (i); the blue circles represent simple oscillation (ii); and the red $\times$s represent the coexistence of a slow cycle and a rapid oscillation (iii). }
    \label{fig:phase}
\end{center}
\end{figure}

We study the sign of the real part of the eigenvalues from eq.(\ref{eq_3eigen}) that determine the stability of the steady state and examine the dependence of the sign on $\beta$ and $\gamma$. Firstly, if $\beta = 0$ and $\gamma \leqslant 0$, there are six negative eigenvalues, where the perfect adaptation state is stable. If $\beta > 0$ or $\beta = 0$ and $\gamma > 0$, there are two positive real eigenvalues and two pairs of complex conjugate eigenvalues with negative real parts. 
If $\beta < 0$, there are four eigenvalues with positive real parts. 
The lattice region in  Fig.\ref{fig:phase} (where either both $\beta$ and $\gamma$ are negative, or even if $\gamma$ is positive, the magnitude of $\beta<0$ is much larger) corresponds to two pairs of complex conjugate eigenvalues with positive real parts. The shaded region in Fig.\ref{fig:phase} corresponds to two real positive eigenvalues and a pair of complex conjugate eigenvalues with positive real parts. 

Interestingly, this classification on the basis of the eigenvalues of the stability matrix corresponds to the classification on the basis of the three behaviors (i) $-$ (iii). We classify the networks that show a particular behavior on the basis of the indices $\beta$ and $\gamma$ and plot each type of behavior in a phase diagram in Fig.\ref{fig:phase}. As shown by the linear stability analysis, perfect adaptation occurs when $\beta = 0$ and $\gamma \leqslant 0$. Simple oscillation appears under the condition $\beta > 0$ or $\beta = 0$ and $\gamma > 0$ and also in the shaded region corresponding to $\beta < 0$ and $\gamma > 0$. The coexistence of rapid burst-type oscillation and the slow cycle occurs at $\beta < 0$, except under the conditions corresponding to the shaded area. 

In other words, phase (ii) appears when the system has no frustration ($\beta > 0$). Because the elements can cooperate even when the original fixed point is unstable, the system belongs to the same category as that for $\alpha >0$ in the two-element case, and thus, a limit cycle is generated. Phase (iii) appears when the system has frustration. There are two pairs of complex conjugate eigenvalues. One pair has a large positive real part and the corresponding eigenvector is orthogonal to the adiabatic space. The real part of the other pair is positive but has a much smaller value; the eigenvector of this pair is not orthogonal to the adiabatic space, and hence, in the adiabatic case, the instability of the fixed point solution is weak, suggesting the emergence of a slow oscillatory mode.
Therefore, the coexistence of the two types of limit cycle attractors, one with a short period and the other with a much longer period, is a natural outcome of the appearance of phase (iii).

In addition to a pair of complex conjugate eigenvalues with a positive real part, the shaded area in Fig.\ref{fig:phase} corresponds to two more positive eigenvalues that are real. Unlike the region for phase (iii), there is no additional complex conjugate pair. In fact, only a single limit cycle arising from the single pair of complex eigenvalues with a real part exists, and consequently, a simple oscillation is developed, as in phase (ii).  The modes with two positive real eigenvalues do not generate a different attractor. Even though three-body frustration exists, as indicated by $\beta <0$, the cooperativity of two-body interactions cancels the frustration. However, there still seems to be a difference between the behavior when $\beta <0$ and the simple oscillation when $\beta >0$. The period in this region is shown in Fig.\ref{fig:Tsys} as the period of network (ii)'. 

\begin{center}
\textbf{V. CASES INVOLVING LARGER NUMBER OF ELEMENTS}
\end{center}

\begin{figure}
\begin{center}
\scalebox{0.52}{\includegraphics{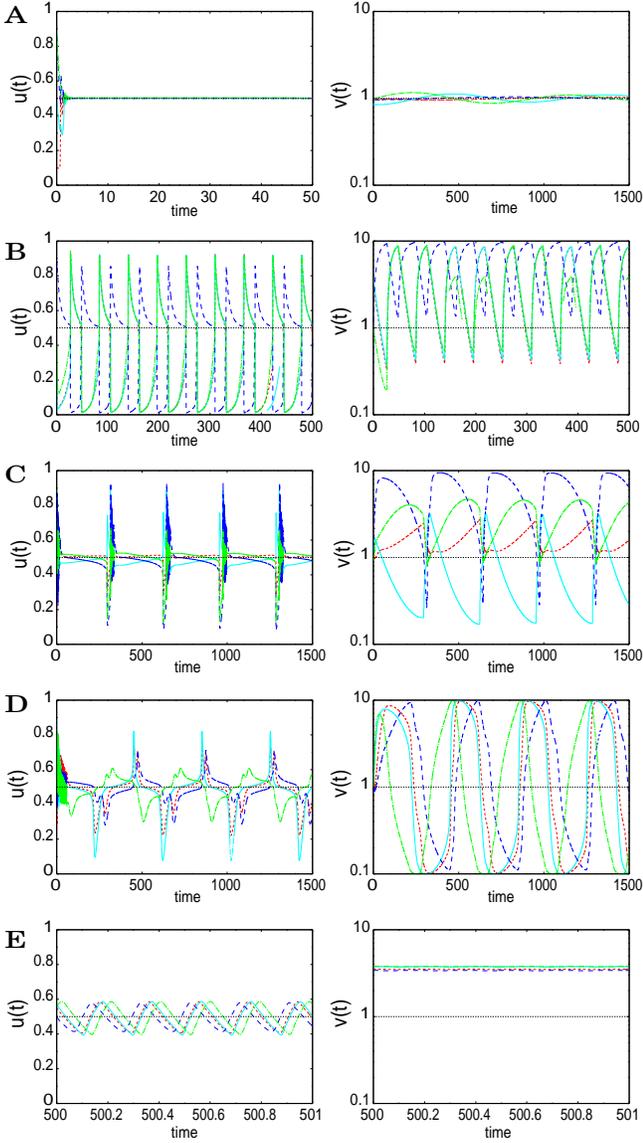}}
\caption{(Color online) Examples of time series of $u$ and $v$ for four-element systems. $\mathbf{A}$ ($C_{12}=0, C_{13}=-1, C_{14}=0, C_{21}=0, C_{23}=-1, C_{24}=-1, C_{31}=1, C_{32}=0, C_{34}=0, C_{41}=-1, C_{42}=1, C_{43}=0$) corresponds to perfect adaptation behavior and $\mathbf{B}$ ($C_{12}=1, C_{13}=-1, C_{14}=1, C_{21}=0, C_{23}=-1, C_{24}=1, C_{31}=-1, C_{32}=0, C_{34}=-1, C_{41}=0, C_{42}=0, C_{43}=-1$) corresponds to oscillatory behavior. $\mathbf{C}$ ($C_{12}=-1, C_{13}=0, C_{14}=-1, C_{21}=-1, C_{23}=1, C_{24}=0, C_{31}=0, C_{32}=-1, C_{34}=0, C_{41}=1, C_{42}=-1, C_{43}=-1$) and $\mathbf{D}$ ($C_{12}=0, C_{13}=1, C_{14}=-1, C_{21}=0, C_{23}=1, C_{24}=0, C_{31}=0, C_{32}=-1, C_{34}=-1, C_{41}=1, C_{42}=1, C_{43}=0$) correspond to slow-cycle behavior and $\mathbf{E}$ (same network as in (d)) corresponds to rapid oscillation. Note carefully the time range in each graph. In each picture, the red dotted line shows the dynamics of element $1$; the cyan solid line, the dynamics of element $2$; the blue dashed line, the dynamics of element $3$; and the light-green chained line, the dynamics of element $4$.
              }
    \label{fig:4dynamics}
\end{center}
\end{figure}

Now, we study a case involving a larger number of elements ($N\geqslant 4$). In such cases, we once again observed three types of behaviors; (i) perfect adaptation, (ii) simple oscillation, and (iii) coexistence of a rapid burst-type oscillation and a slow cycle (Fig.\ref{fig:4dynamics}). Thus far, we have not observed behaviors that deviate from the behaviors in these cases. 

To classify the network behaviors, we again performed linear stability analysis of eq.(\ref{eq_adapt_re}) around the original, perfectly adapted fixed point solution. In this case, the dominant structures in the interaction network are again characterized by the coefficients of the characteristic equation, which indicate the $n$-loop frustration in the network. 
The coefficients are given by loop structures over $\textit{n}$ elements ($\textit{n} = 2,3, \cdots ,N$) containing a product of smaller loops as follows:
\begin{align}
\mathcal{L} (2) &= (\frac{\kappa \sigma}{2} )^2 \sum_{<ij>} C_{ij} C_{ji} , \\
\mathcal{L} (3) &= (\frac{\kappa \sigma}{2} )^3 \sum_{<ijk>} C_{ij} C_{jk} C_{ki} , \\
\mathcal{L} (4) &= (\frac{\kappa \sigma}{2} )^4 (\sum_{<ijkl>} C_{ij}C_{jk}C_{kl}C_{li}  \nonumber  \\ 
 & \ \ \ \ \ \ \ \ \ \ \ \ \ \ \ \ \ \ \   - \sum_{<ij><kl>} C_{ij}C_{ji} \times C_{kl}C_{lk}), 
 \label{eq_L4}
 \end{align}
where the summation by $<ij>$ runs over all possible pairs with $i \neq j$ and that by $<ijk>$ runs over all triplets with $i \neq j$, $j \neq k$ and $k \neq i$; the pairs and triplets are chosen from among the $N$ elements.

For the case in which more than four elements are involved, the $n$-loop frustration is characterized by the $\mathcal{L} (n)$ that is determined by a product of $\mathcal{L} (m_j) \ (m_j \geqslant 2, \ \sum_{j=1}^{M} m_j = n)$, as follows: 
\begin{align}
\mathcal{L} (n) &=  \sum_{<i_1\cdots i_n>} C_{i_1 i_2} C_{i_2 i_3} \cdots C_{i_n i_1}  \nonumber  \\ 
&+ \sum_{m_1} \cdots \sum_{m_M} \frac{(-1)^{m_1 +1} \cdots (-1)^{m_M +1}}{(-1)^{n+1}} \mathcal{L} (m_1) \cdots \mathcal{L} (m_M) .
 \label{eq_Ln}
\end{align}

\begin{figure}
\begin{center}
\scalebox{0.4}{\includegraphics{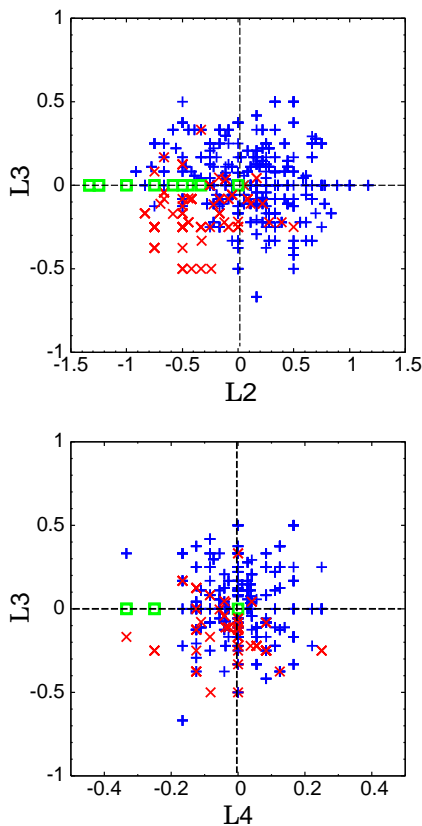}}
\caption{(Color online) Classification of four-element systems on the basis of $\mathcal{L} (2)$(abscissa; upper figure) and $\mathcal{L} (3)$(ordinate; upper figure) as well as $\mathcal{L} (4)$(abscissa; lower figure) and $\mathcal{L} (3)$(ordinate; lower figure). The indices are defined in the text. The green box represents a network that exhibits perfect adaptation behavior; the blue $+$s represent  simple oscillation; the red $\times$s represent the coexistence of a slow cycle and a rapid oscillation. $1,000$ networks are generated randomly and each network is labeled as a particular type on the basis of the period $T_{sys}$.
              }
    \label{fig:4phase}
\end{center}
\end{figure}

For $N=4$, we again draw a phase diagram for the three types of behaviors, which are plotted with respect to the three $\mathcal{L} $ indices, by classifying the networks on the basis of the period of the attractor(s). 
The classification of the networks on the basis of the sign of $\mathcal{L} $ as in the case where $N=3$ seems to be valid for the case where $N=4$ as well (Fig.\ref{fig:4phase}).

\begin{center}
\textbf{VI. DISCUSSION AND CONCLUSION}
\end{center}

In this study, we have introduced a system of coupled adaptive elements, studied the behavior of the dynamics of these elements, and classified the behavior on the basis of its relationship with the network structure. For a system with two elements, two types of behaviors, perfect adaptation and simple oscillation, were observed. For a system with three elements, novel types of dynamics, namely, rapid burst-type oscillation and a slow cycle, were discovered in addition to the above two behaviors. The rapid oscillation and the slow cycle may coexist depending on the initial conditions. These dynamics are a result of the existence of two distinct time scales for the behavior of the adaptive element, i.e., fast response and slow adaptation processes; hence, the dynamics are characteristic of coupled adaptive systems. The dynamics of each network are classified on the basis of the network structure (in specific, a combination of positive or negative feedback among elements). Cooperativity among elements leads to simple oscillation, whereas frustration among elements leads to the coexistence of a rapid bursting oscillation and a slow cycle. The index for frustration is defined by the product of signs of the interaction matrix elements over the entire loop. This index also characterizes the eigenvalues in the linear stability matrix around the steady state. 

The period of the rapid burst-type oscillation is on the order of the response time in a single element, implying the total absence of adaptation; in contrast, the period of the slow cycle is much larger than the original time scale of adaptation, where frustration among elements caused by interaction was important. It is remarkable that the attractors with such large differences in their periods generally coexist.
Frustration in a coupled dynamical system is also discussed in \cite{Bersini_JTB97}, where frustration leads to a chaotic behavior. 

Coupled oscillators have attracted much attention and many general concepts as synchronization have been developed\cite{Kuramoto}. In oscillator networks, frustration causes synchronization\cite{Zenette_EL05} or leads to an ordered state characterized by quasientrainment\cite{Daido_PRL92}. Synchronization in coupled chaotic oscillators have also been studied\cite{Pykovsky_PRL96, Pykovsky_PRL97}. However, a coupled system of adaptive elements has not been studied thus far. Our study here demonstrates that novel cooperative dynamics generally arise as a result of frustration in coupled adaptive systems.

Differences in the response time scales are commonly observed in biological systems. The frequency of the calcium oscillations in a cell often depends on the cellular state, and it is proposed that this frequency is related to cell type differenitaion\cite{Dolmetsch_Nature98}. Neuroscience is another example, where two-state fluctuations in neural activity have recently attracted much attention\cite{Wilson_JN96, Stern_JN97, Anderson_Nature00}; where a neuron spontaneously switches between the up state with rapid-burst firing and down state with rare firing. Some models have been proposed to explain the mechanics of two-state fluctuations by taking into account the recursive excitable interaction\cite{Compte_JN03, Kang_NN04}; however, the mechanism is not yet fully understood. Although our model has not been developed specifically for the neural system, it shows excitatory response that is common in the neural system. Our assertion in this context is simple: frustration in interactions among adaptive elements generates bistability between burst firing and rare firing.

The extension of the three-element system to a larger number of elements is straightforward. Once again, we observed the coexistence of burst-type oscillations and a slow cycle. The behaviors are classified on the basis of the frustration in a loop structure over $\textit{n}$ elements ($\textit{n} = 2,3, \cdots ,N$) in an interaction network; the frustration in the interaction network is characterized by the coefficients in a characteristic equation in the linear stability analysis around the steady state. These frustration indices are again relevant to the classification. 

The generation of long-term dynamics and differences in time scales is essential in determining cellular and neural behaviors. Studies focusing on the time-scale difference, a coupled chaotic system with a variety of time scales\cite{Fujimoto_Chaos03} or coupled phase-oscillators with diversification of the time scales\cite{Armbruster_PL99, Fujimoto_EPL04} have been carried out. 
Depending on the network structure, such behavior is inherent to a coupled adaptive system in a network, despite the simplicity of the model that is utilized in this case. Here, we have only studied the case in which the coupling between identical elements is the same. 
In biology, elements are often heterogeneous and the number of elements can be very large. Such complex cases are currently being investigated, and these investigations will be reported elsewhere. 

The authors would like to thank A.Awazu, S.Ishihara, K. Fujimoto, and D.Shimaoka for their valuable comments.
MI was partially supported by the JSPS Research Fellowships for Young Scientists.

\end{document}